\documentclass[aps,prb,twocolumn,superscriptaddress,showpacs,showkeys,footinbib]{revtex4}

\usepackage{amsmath}
\usepackage{amssymb}
\usepackage{graphicx}
\usepackage{relsize}
\usepackage{bm}
\usepackage{color}

\renewcommand{\i}{\ensuremath{\mathrm{i}}}
\newcommand{\e}{\ensuremath{\mathrm{e}}}
\renewcommand{\d}{\ensuremath{\mathrm{d}}}

\begin{document}
\title{Magneto-optical conductivity of graphene on polar substrates}
\author{Benedikt Scharf}
\affiliation{Department of Physics, University at Buffalo, State University of New York, Buffalo, NY 14260, USA}
\author{Vasili Perebeinos}
\affiliation{IBM T. J. Watson Research Center, Yorktown Heights, NY 10598, USA}
\author{Jaroslav Fabian}
\affiliation{Institute for Theoretical Physics, University of Regensburg, 93040 Regensburg, Germany}
\author{Igor \v{Z}uti\'c}
\affiliation{Department of Physics, University at Buffalo, State University of New York, Buffalo, NY 14260, USA}

\date{\today}

\begin{abstract}
We theoretically study the effect of polar substrates on the magneto-optical conductivity of doped monolayer graphene, where we particularly focus on the role played by surface polar phonons (SPPs). Our calculations suggest that polaronic shifts of the intra- and interband absorption peaks can be significantly larger for substrates with strong electron-SPP coupling than those in graphene on non-polar substrates, where only intrinsic graphene optical phonons with much higher energies contribute. Electron-phonon scattering and phonon-assisted transitions are, moreover, found to result in a loss of spectral weight at the absorption peaks. The strength of these processes is strongly temperature-dependent and with increasing temperatures the magneto-optical conductivity becomes increasingly affected by polar substrates, most noticeably in polar substrates with small SPP energies such as HfO$_2$. The inclusion of a Landau level-dependent scattering rate to account for Coulomb impurity scattering does not alter this qualitative picture, but can play an important role in determining the lineshape of the absorption peaks, especially at low temperatures, where impurity scattering dominates.
\end{abstract}


\maketitle

\section{Introduction}\label{Sec:Intro}

\begin{figure}[t]
\centering
\includegraphics*[width=8cm]{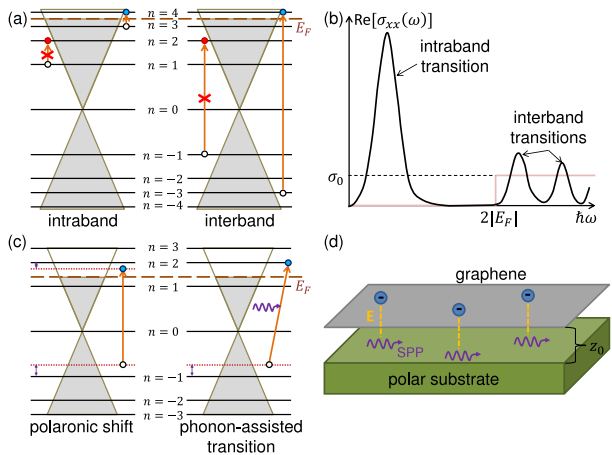}
\caption{(Color online) (a) Optical intra- and interband transitions between Landau levels (LLs) $n$ at $T=0$ (in the absence of Zeeman splitting). The transitions are governed by the selection rule $|n|\to|n\pm1|$, must conserve momentum and the electron spin, and are permitted only if the initial state (white circle) is occupied and the final state (blue or red circles) is unoccupied. (b) Sketch of the magneto-optical conductivity. (c) The electron-phonon coupling leads to the emergence of dressed LLs, which are shifted (or even split) with respect to the bare LLs and between which optical transitions then occur. Moreover, phonon-assisted transitions, that is, momentum-conserving transitions involving the absorption or emission of a phonon, can occur. (d) Surface polar phonons (SPPs) couple to electrons in graphene via the electric near-field caused by the SPPs.}\label{fig:Scheme}
\end{figure}

One of the main reasons for the tremendous interest shown in graphene during the past decade is that its excellent transport and optical properties\cite{CastroNeto2009:RMP,Peres2010:RMP,DasSarma2011:RMP} make it an attractive candidate for possible applications in nanoscale electronics and optoelectronics.\cite{Geim2009:Science,Avouris2010:NL,Ferrari2010:NP} Recent breakthroughs\cite{Tombros2007:Nature,Cho2007:APL,Nishioka2007:APL,Han2009:PRL,Han2010:PRL,McCreary2012:PRL,Gmitra2013:PRL,Neumann2013:unpublished} in graphene indicate that it may also be particularly suitable for spintronics.\cite{Zutic2004:RMP,Fabian2007:APS,Josza2011:HSTM,Dery2012:IEEE,Seneor2012:MRS}

In addition to its possible technological applications, graphene is also highly appealing for fundamental research: Its low-energy excitations can be described by a 2D Dirac-like Hamiltonian of massless fermions with an effective speed of light $v_\mathrm{\mathsmaller{F}}\approx 10^8$ cm/s, which essentially allows one to study quantum electrodynamics in (2+1) dimensions by studying the electronic properties of graphene.\cite{CastroNeto2009:RMP} The Dirac-like behavior of electrons near the K and K' points in graphene is also reflected in their response to a magnetic field. If a perpendicular magnetic field is applied to the graphene plane, the linear dispersion in the vicinity of those points evolves into a discrete spectrum of non-equidistant Landau levels (LLs), which gives rise to an unconventional, half-integer quantum Hall effect.\cite{Gusynin2005:PRL,Peres2006:PRB}

Likewise, graphene subject to a magnetic field exhibits peculiar optical properties, conveniently described by the so-called magneto-optical conductivity, the optical conductivity in the presence of a magnetic field: In contrast to conventional two-dimensional electron gases, which only exhibit one pronounced optical absorption peak centered around the cyclotron resonance frequency, a sequence of distinct optical absorption peaks can be experimentally observed in monolayer graphene\cite{Sadowski2006:PRL,Jiang2007:PRL,Sadowski2007:SolidStateComm,Deacon2007:PRB,Henriksen2010:PRL,Orlita2010:SST,Crassee2011:PRB,Orlita2011:PRL,Ellis2012:IRMMW} due to its non-equidistant LL spectrum.\cite{Gusynin2007:PRL,Koshino2008:PRB} Those peaks broadly correspond to optical transitions between different LLs $n$, which must satisfy Pauli's principle [see Figs.~\ref{fig:Scheme}~(a) and~(b)].

The interaction between electrons and phonons modifies this simple single-particle picture in the following way [see Fig.~\ref{fig:Scheme}~(c)]: (i) Electrons moving through the crystal lattice generate a polarization field, which results in the emergence of polarons, quasiparticles describing an electron and its accompanying polarization field. The formation of polarons modifies the individual LLs giving rise to dressed (and even split) LLs, which in turn leads to shifts of the absorption peaks observed in the magneto-optical conductivity. (ii) During optical transitions phonons can be absorbed or emitted, which results in the appearance of phonon sidepeaks in the absorption spectrum. Both effects have been discussed in impressive detail in Ref.~\onlinecite{Pound2012:PRB} for the case of graphene. In addition, the formation of new electron-phonon bound states---both at zero\cite{Badalyan2012:PRB} and finite\cite{Zhu2012:PRL} magnetic field---further refines this picture.

If graphene is situated on a polar substrate, electrons couple not only to intrinsic graphene phonons, but also to surface polar phonons (SPPs), that is, surface phonons of polar substrates which interact with the electrons in graphene via the electric fields those phonons generate [see Fig.~\ref{fig:Scheme}~(d)]. Those phonons can have a profound impact on transport or optical properties of graphene: Surface polar phonons have, for example, been argued to be responsible for current saturation in graphene\cite{Meric2008:NatureNano,Freitag2009:NL,Perebeinos2010:PRB,Konar2010:PRB,Li2010:APL,DaSilva2010:PRL} and to affect the carrier mobility,\cite{Chen2008:NatureNano,Fratini2008:PRB,Zhang2013:PRB} which can also be influenced by the emergence of interfacial plasmon-phonon modes due to the coupling between SPPs and graphene plasmons,\cite{Ong2012:PRB,Ong2013:PRB} the spin-relaxation,\cite{Ertler2009:PRB} the optical absorption,\cite{Scharf2013:PRB} the renormalization of the Fermi velocity\cite{Hwang2013:PRB} as well as the plasmon dispersion and damping in graphene.\cite{Yan2013:NP} Thus, it stands to reason that magneto-optical experiments with graphene on polar substrates would likewise be affected by SPPs.

Since SPPs break the space inversion symmetry, they lead to an effective Bychkov-Rashba spin-orbit coupling and another channel for spin relaxation in graphene.\cite{Ertler2009:PRB} A detailed understanding of SPPs in graphene and the related magneto-optical experiments can thus also provide important insights for graphene spintronics, especially in the regime where other contributions to spin relaxation could be minimized.

Our main goal in this manuscript is to study the role played by SPPs (and---by extension---the substrate) and how they affect the magneto-optical conductivity. To do so, we use linear response theory to derive a Kubo formula for the magneto-optical conductivity in Sec.~\ref{Sec:Model}, which is then evaluated for monolayer graphene on several different (polar) substrates in Sec.~\ref{Sec:Results}. We find that the polaronic shift of the absorption peaks strongly depends on the substrate and can be significantly larger in polar substrates as compared to suspended graphene or graphene on a non-polar substrate such as diamond-like carbon. Moreover, polar substrates introduce a strong temperature dependence of the inelastic electron-phonon scattering as well as of the polaronic shifts.

\section{Model and formalism}\label{Sec:Model}
In this work we consider a graphene monolayer situated on a polar substrate. Using the Dirac-cone approximation, the electronic single-particle spectrum of monolayer graphene in the presence of a magnetic field of magnitude $B$ and perpendicular to the graphene plane ($xy$-plane) reads as
\begin{equation}\label{electronic_single_particle_spectrum}
\epsilon_{s}(n)=\frac{\mathrm{sgn}(n)\sqrt{2|n|}\hbar v_\mathrm{\mathsmaller{F}}}{l_\mathsmaller{B}}+s\frac{g\mu_\mathsmaller{B}B}{2},
\end{equation}
with the magnetic length $l_\mathsmaller{B}=\sqrt{\hbar/(eB)}$, the Bohr magneton $\mu_\mathsmaller{B}$, the $g$-factor $g=2$, and the absolute value of the electron charge $e=|e|$. Here $n\in\mathbb{Z}$ denotes a LL in the conduction ($n\geq0$) or valence bands ($n\leq0$), while $s$ denotes the spin-degree of freedom along the $z$-direction.\cite{McClure1956:PhysRev,CastroNeto2009:RMP} The spectrum~(\ref{electronic_single_particle_spectrum}) is strongly degenerate as each single-electron state is specified by the quantum numbers $n$ and $s$ as well as the valley and momentum quantum numbers $v$ and $k$. Introducing the corresponding creation and annihilation operators $\hat{c}^{\dagger}_{nksv}$ and $\hat{c}_{nksv}$, we can write the electronic single-particle Hamiltonian as
\begin{equation}\label{Hamiltonian_electrons}
\hat{H}_{\mathrm{e}}=\sum\limits_{n,k,s,v,\lambda}\epsilon_{s}(n)\,\hat{c}^\dagger_{nksv}\hat{c}_{nksv}.
\end{equation}

In addition our model also includes different phonon branches (labeled as $\Lambda$). Those phonons are described by their dispersion $\omega_{\mathsmaller{\Lambda}}\left(\mathbf{q}\right)$ and momentum $\mathbf{q}$, their corresponding creation (annihilation) operators $\hat{p}^\dagger_{\mathbf{q}\Lambda}$ $(\hat{p}_{\mathbf{q}\Lambda})$, and the Hamiltonian
\begin{equation}\label{Hamiltonian_phonons}
\hat{H}_{\mathrm{ph}}=\sum\limits_{\mathbf{q},\Lambda}\hbar\omega_{\mathsmaller{\Lambda}}\left(\mathbf{q}\right)\hat{p}^\dagger_{\mathbf{q}\Lambda}\hat{p}_{\mathbf{q}\Lambda}.
\end{equation}
Moreover, the electrons and phonons interact with each other and the coupling between both systems is given by
\begin{equation}\label{Hamiltonian_coupling}
\begin{aligned}
\hat{H}_{\mathrm{e-ph}}=\sum\limits_{nksv}\sum\limits_{n'\mathbf{q}v'\Lambda}&\alpha_{vv',\Lambda}^{nn'}(k,\mathbf{q})\left(\hat{p}^\dagger_{-\mathbf{q}\Lambda}+\hat{p}_{\mathbf{q}\Lambda}\right)\\
&\times\hat{c}^\dagger_{n(k+q_x)sv}\hat{c}_{n'ksv'},
\end{aligned}
\end{equation}
where $\alpha_{vv',\Lambda}^{nn'}(k,\mathbf{q})$ is the electron-phonon coupling matrix element. The relevant phonons considered here can be divided into two different categories: intrinsic graphene optical phonons and, if graphene is situated on a polar substrate, SPPs. If graphene is situated on a non-polar substrate, only intrinsic graphene phonons are included. Thus, the total Hamiltonian of our model reads as
\begin{equation}\label{Hamiltonian}
\hat{H}=\hat{H}_{\mathrm{e}}+\hat{H}_{\mathrm{ph}}+\hat{H}_{\mathrm{e-ph}}.
\end{equation}

\begin{figure}[t]
\centering
\includegraphics*[width=8cm]{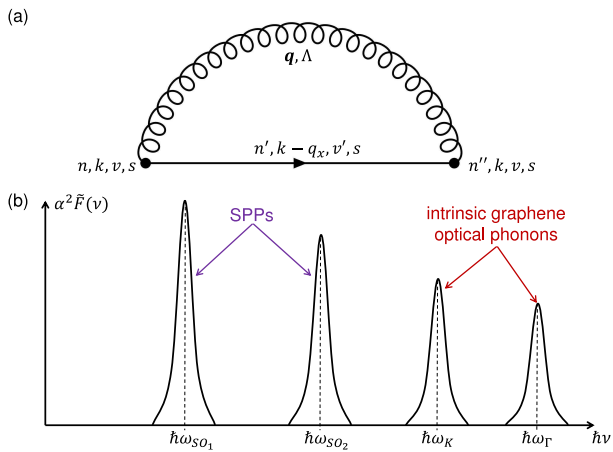}
\caption{(Color online) (a) Feynman diagram corresponding to the lowest-order self-energy. Solid and wiggly lines denote electron and phonon propagators, respectively. Equation~(\ref{self_energy_phonons}) is obtained by averaging over the electron-phonon coupling matrix elements. (b) Schematic view of the frequency dependence of the phonon spectral function $\alpha^2\tilde{F}(\nu)$ for a typical polar substrate.}\label{fig:Feynman}
\end{figure}

If the electron-phonon interaction is weak, the coupling can be treated using standard diagrammatic perturbation theory.\cite{Mahan2000} Due to the interaction with phonons the energy of a single electron is modified, where the additional contribution to the energy arising from this interaction is described by the self-energy. For an electron with spin quantum number $s$ the retarded electronic self-energy due to the electron-phonon coupling can---in the lowest order [see Fig.~\ref{fig:Feynman}~(a)]---be approximated by
\begin{equation}\label{self_energy_phonons}
\begin{aligned}
\Sigma&^{\mathsmaller{\mathrm{R}}}_{\mathsmaller{\mathrm{ph},s}}\left(\omega\right)=\int\limits_0^\infty\d\nu\alpha^2\tilde{F}(\nu)\int\limits_{-\infty}^\infty\d\omega'N_s(\hbar\omega'+\mu)\\
&\times\bigg[\frac{n_\mathsmaller{\mathrm{BE}}\left(\hbar\nu\right)+n_\mathsmaller{\mathrm{FD}}\left(-\hbar\omega'\right)}{\omega-\omega'-\nu+\i0^\mathsmaller{+}}+\frac{n_\mathsmaller{\mathrm{BE}}\left(\hbar\nu\right)+n_\mathsmaller{\mathrm{FD}}\left(\hbar\omega'\right)}{\omega-\omega'+\nu+\i0^\mathsmaller{+}}\bigg]
\end{aligned}
\end{equation}
following Migdal's approach and averaging over the electron-phonon coupling matrix elements at the Fermi surface.\cite{Migdal1958:JETP,Allen1982:SSP,Dogan2003:PRB,Nicol2009:PRB,Pound2011:PRB,Pound2012:PRB} Here $\alpha^2\tilde{F}(\nu)$ denotes the Eliashberg electron-phonon spectral function normalized to the electronic density of states (DOS) per spin at the chemical potential $\mu=\mu(T)$ and $N_s(\epsilon)$ the electronic DOS for spin $s$, while $n_\mathsmaller{\mathrm{FD/BE}}(\epsilon)=1/[\exp(\beta\epsilon)\pm1]$, with $\beta=1/(k_{\mathsmaller{\mathrm{B}}}T)$, the temperature $T$, and the Boltzmann constant $k_{\mathsmaller{\mathrm{B}}}$, are the Fermi-Dirac and Bose-Einstein distribution functions, respectively.\footnote{Both Eq.~(\ref{self_energy_phonons}) and averaging the phonon spectral function and electron-phonon coupling matrix elements over the Fermi surface constitute very good approximations to calculate the electronic self-energy if the electron-phonon coupling or the typical phonon energies compared to the Fermi energy are small (see Refs.~\onlinecite{Allen1982:SSP,Dogan2003:PRB}).}

The spectral function $\alpha^2\tilde{F}(\nu)$ contains the averaged (and spin-independent) coupling of electrons in graphene to the optical and SPPs. In the following, we will assume that the phonon frequencies and the electron-phonon coupling are not significantly affected by the magnetic field and that the averaged coupling can be reasonably well described by its zero-field value. Then the dominant electron-optical phonon coupling is that to longitudinal-optical (LO) and transverse-optical (TO) phonons at the $\Gamma$ point and to the TO phonon at the $K$ point.\cite{Piscanec2004:PRL,Lazzeri2005:PRL,Perebeinos2005:PRL2,Ando2006:JPSJ,Lazzeri2008:PRB,Borysenko2010:PRB} The dispersion of LO and TO phonons near the $\Gamma$ point can be approximated by the constant energy $\hbar\omega_\mathsmaller{\Gamma}\approx197$ meV, that of TO phonons near the $K$ and $K'$ points by $\hbar\omega_\mathsmaller{K}\approx157$ meV. Moreover, there are typically two surface optical (SO) phonons in polar substrates that interact with the electrons in graphene and whose dispersion can again be approximated by substrate-specific, constant frequencies $\omega_\mathsmaller{\mathrm{SO}_1}$ and $\omega_\mathsmaller{\mathrm{SO}_2}$ summarized in Table~\ref{tab:parameters}.

The resulting phonon spectral function in our model then reads as
\begin{equation}\label{phonon_spectral_function}
\alpha^2\tilde{F}(\nu)=\sum\limits_\mathsmaller{\Lambda} A_\Lambda\delta\left(\nu-\omega_\mathsmaller{\Lambda}\right),
\end{equation}
which consists of four contributions [see Fig.~\ref{fig:Feynman}~(b)]: the combined contribution from both $\Gamma$ phonon modes with a total coupling\cite{Piscanec2004:PRL,Lazzeri2005:PRL,Perebeinos2005:PRL2,Ando2006:JPSJ,Lazzeri2008:PRB,Borysenko2010:PRB} $A_\mathsmaller{\Gamma}=\hbar D^2_\mathsmaller{\Gamma}A/(2M_\mathsmaller{\mathrm{c}}\omega_\mathsmaller{\Gamma})$, the contribution from  the $K_\mathsmaller{\mathrm{TO}}$ phonon mode with a coupling twice as large as that of each phonon at the $\Gamma$ point,\cite{Piscanec2004:PRL,Lazzeri2005:PRL,Perebeinos2005:PRL2} $A_\mathsmaller{K}=\hbar D^2_\mathsmaller{\Gamma}A/(2M_\mathsmaller{\mathrm{c}}\omega_\mathsmaller{K})$, and the contribution from the $\Lambda=\mathrm{SO}_1$ and $\Lambda=\mathrm{SO}_2$ modes described by the coupling\cite{Fratini2008:PRB,Konar2010:PRB} $A_\mathsmaller{\Lambda}=(e^2\pi/2)\int_0^{2\pi}\d\theta F_{\mathsmaller{\Lambda}}^2[q(\theta)]\,\e^{-2q(\theta)z_0}\left(1+\cos\theta\right)/q(\theta)$ with $q(\theta)\equiv|\mu|\sqrt{2-2\cos\theta}/(\hbar v_\mathrm{\mathsmaller{F}})$. Here $M_\mathsmaller{\mathrm{c}}$ is the carbon mass, $D_\mathsmaller{\Gamma}\approx11.2$ eV/{\AA} the strength of the electron-$\Gamma$ phonon coupling,\cite{Perebeinos2010:PRB} $A=3\sqrt{3}a^2/2$ the area of the graphene unit cell, $a\approx1.42$ {\AA} the distance between two carbon atoms, $z_0\approx3.5$ {\AA} the van der Waals distance between the graphene sheet and the substrate, and $F_{\mathsmaller{\Lambda}}^2(q)$ the Fr\"{o}hlich coupling given by\cite{Wang1972:PRB,Fischetti2001:JoAP,Price2012:PRB,Scharf2013:PRB}
\begin{equation}\label{Froehlich_SOi}
F_{\mathsmaller{\mathrm{SO}_{1/2}}}^2(q)=\frac{\hbar\omega_\mathsmaller{\mathrm{SO}_{1/2}}}{2\pi}\left[\frac{1}{\varepsilon_\mathsmaller{\mathrm{i}/\infty}+\varepsilon(q)}-\frac{1}{\varepsilon_\mathsmaller{0/\mathrm{i}}+\varepsilon(q)}\right]
\end{equation}
with the optical, intermediate, and static permittivities $\varepsilon_\mathsmaller{\infty}$, $\varepsilon_\mathsmaller{\mathrm{i}}$, and $\varepsilon_\mathsmaller{0}$ of the substrate as well as the static, low temperature dielectric function $\varepsilon(q)=1+2\pi e^2\Pi_{\mathsmaller{\mathrm{g}}}\left(q,\omega=0\right)/(\kappa q)$, where $\kappa$ is the background dielectric constant and $\Pi_{\mathsmaller{\mathrm{g}}}\left(q,\omega\right)$ the polarization function of graphene. We find that, for the relatively high doping and the magnetic fields considered in this work, the results are not noticeably affected by whether the polarization function calculated for zero\cite{Wunsch2006:NJoP,Hwang2007:PRB} or finite\cite{Pyatkovskiy2011:PRB} magnetic field is used. In our model the dielectric medium above the graphene plane is air, and for simplicity we use the average $\kappa=(1+\varepsilon_\mathsmaller{0})/2$ for the background dielectric constant. The dielectric function $\varepsilon(q)$ accounts for the screening of the Coulomb interaction in the graphene sheet above the polar substrate.

\begin{table}
\begin{center}
\begin{tabular}{|c||c|c|c|c|c|}
\hline
 & Al$_2$O$_3$$^a$ & h-BN$^b$ & HfO$_2$$^c$ & SiC$^d$ & SiO$_2$$^e$\\
\hline\hline
$\varepsilon_\mathsmaller{0}$ & 12.53 & 5.09 & 22.0 & 9.7 & 3.90\\
\hline
$\varepsilon_\mathsmaller{\mathrm{i}}$ & 7.27 & 4.575 & 6.58 & - & 3.36\\
\hline
$\varepsilon_\mathsmaller{\infty}$ & 3.20 & 4.10 & 5.03 & 6.5 & 2.40\\
\hline
$\hbar\omega_\mathsmaller{\mathrm{SO}_1}$ [meV] & 56.1 & 101.7 & 21.6 & 116.0 & 58.9\\
\hline
$\hbar\omega_\mathsmaller{\mathrm{SO}_2}$ [meV] & 110.1 & 195.7 & 54.2 & - & 156.4\\
\hline
$F^2_{\mathsmaller{\mathrm{SO}_1}}$ [meV] & \; 0.420\;& \; 0.258\;& \; 0.304\;& \; 0.735\;& \; 0.237\;\\
\hline
$F^2_{\mathsmaller{\mathrm{SO}_2}}$ [meV] & \; 2.053\;& \; 0.520\;& \; 0.293\;& \;-\;& \; 1.612\;\\
\hline
\end{tabular}
\end{center}
\caption{Optical, intermediate, and static permittivities as well as frequencies and the strengths of the bare Fr\"{o}hlich couplings [$\varepsilon(q)$ set to 1 in Eq.~(\ref{Froehlich_SOi})] for the SPP scattering on the substrates Al$_2$O$_3$, hexagonal BN, HfO$_2$ SiC, and SiO$_2$ (taken from Ref.~\onlinecite{Scharf2013:PRB}).\\
$^a$ Refs.~\onlinecite{Fischetti2001:JoAP,Ong2012:PRB}, $^b$ Refs.~\onlinecite{Geick1966:PR,Perebeinos2010:PRB,Ong2012:PRB}, $^c$ Refs.~\onlinecite{Fischetti2001:JoAP,Perebeinos2010:PRB,Ong2012:PRB}, $^d$ Refs.~\onlinecite{Harris1995,Perebeinos2010:PRB}, $^e$ Ref.~\onlinecite{Perebeinos2010:PRB}, which uses averages of values from Refs.~\onlinecite{Fischetti2001:JoAP,Fratini2008:PRB,Konar2010:PRB}.}\label{tab:parameters}
\end{table}

Moreover, the spin-polarized DOS for the spectrum given in Eq.~(\ref{electronic_single_particle_spectrum}) is needed to calculate the self-energy~(\ref{self_energy_phonons}), and we also include the effect of scattering (other than scattering with optical or SPPs, such as scattering at charged impurities) on a phenomenological level by introducing the constant scattering rate $\Gamma$. The self-energy obtained in this way is then inserted into Kubo formulas for the magneto-optical conductivities $\sigma_{xx}(\omega)=\sigma_{yy}(\omega)$ and $\sigma_{xy}(\omega)=-\sigma_{yx}(\omega)$, and we refer to the Appendix~\ref{Sec:Appendix} for further details on this procedure. Here we are primarily interested in the absorption, that is, essentially in $\mathrm{Re}\left[\sigma_\mathsmaller{xx}(\omega)\right]$. In the following, we will thus compute the real part of $\sigma_{xx}(\omega)$ as well as the imaginary part of $\sigma_{xy}(\omega)$ following the procedure outlined in the Appendix~\ref{Sec:Appendix}. Those calculations are performed for several different substrates with the corresponding parameters summarized in Table~\ref{tab:parameters} and the results discussed in the following section.

\section{Results}\label{Sec:Results}

\begin{figure}[t]
\centering
\includegraphics*[width=8cm]{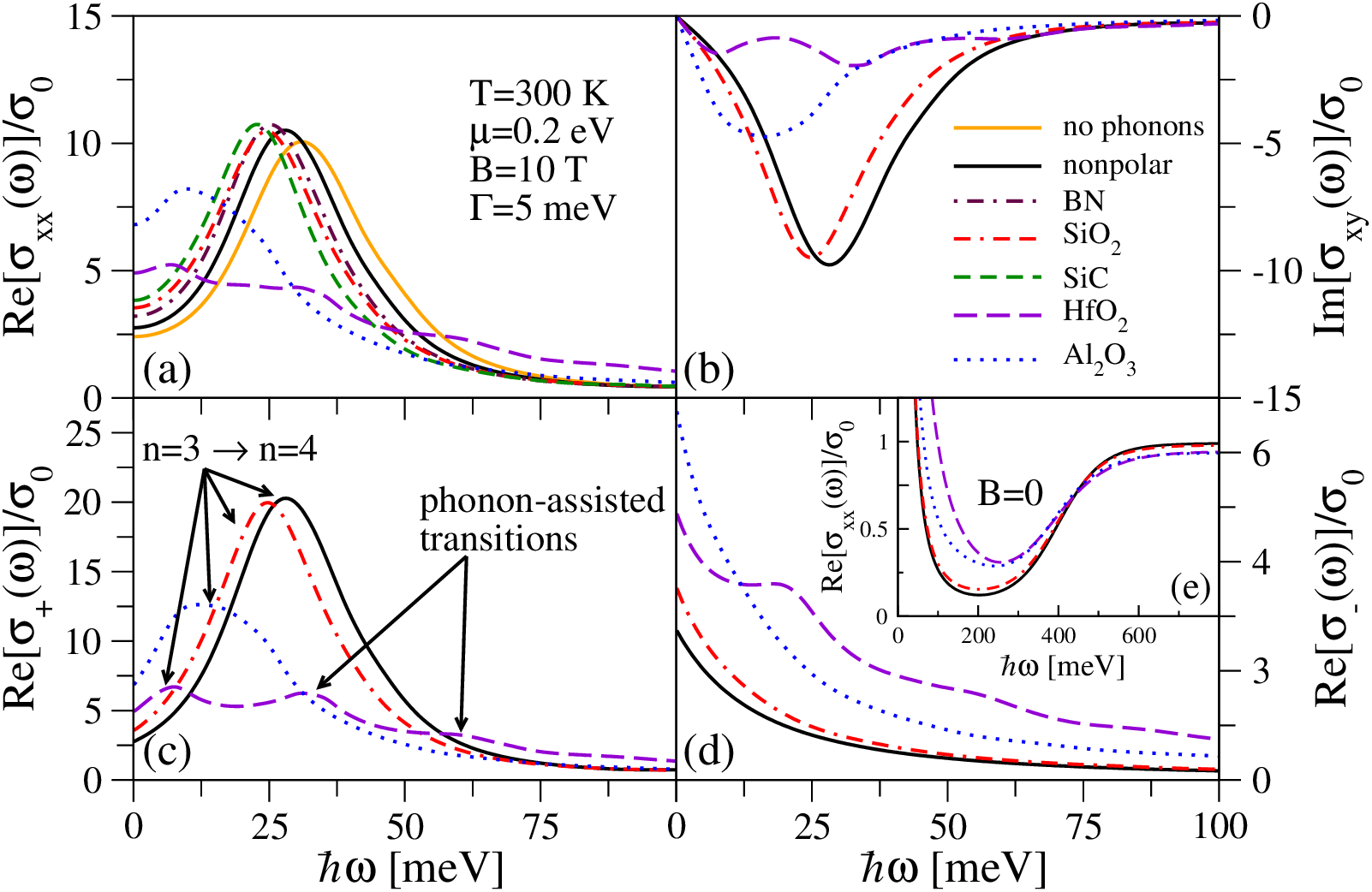}
\caption{(Color online) Calculated frequency dependence of the magneto-optical conductivity of graphene on several different substrates at room temperature: (a) $\mathrm{Re}\left[\sigma_\mathsmaller{xx}(\omega)\right]$,  (b) $\mathrm{Im}\left[\sigma_\mathsmaller{xy}(\omega)\right]$, (c) $\mathrm{Re}\left[\sigma_\mathsmaller{+}(\omega)\right]$, (d) $\mathrm{Re}\left[\sigma_\mathsmaller{-}(\omega)\right]$. For comparison, $\mathrm{Re}\left[\sigma_\mathsmaller{xx}(\omega)\right]$ for zero magnetic field is presented in (e).
}\label{fig:GeneralConductivity}
\end{figure}

\begin{figure}[t]
\centering
\includegraphics*[width=8cm]{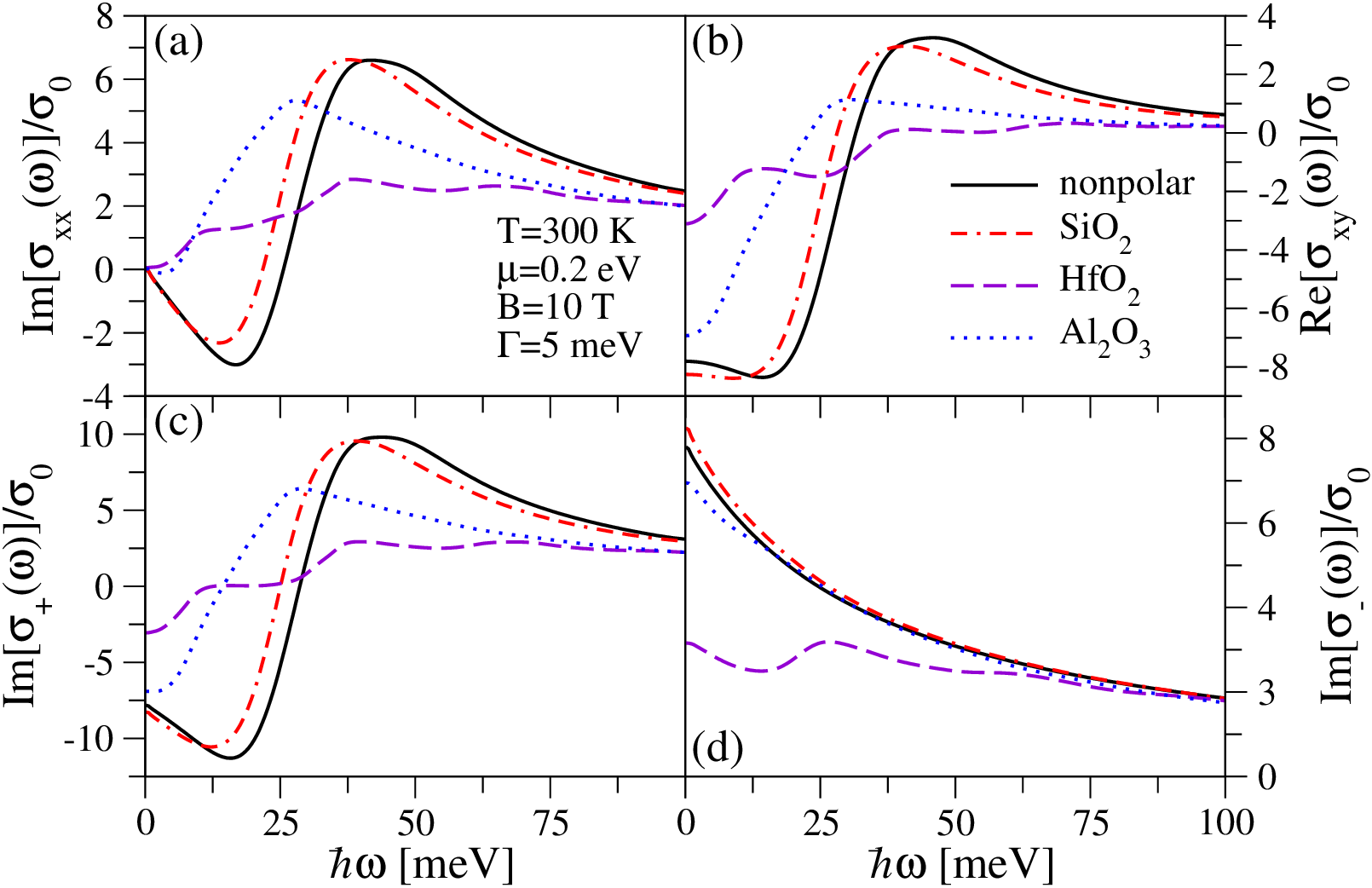}
\caption{(Color online) Calculated frequency dependence of the magneto-optical conductivity of graphene on several different substrates at room temperature: (a) $\mathrm{Im}\left[\sigma_\mathsmaller{xx}(\omega)\right]$,  (b) $\mathrm{Re}\left[\sigma_\mathsmaller{xy}(\omega)\right]$, (c) $\mathrm{Im}\left[\sigma_\mathsmaller{+}(\omega)\right]$, (d) $\mathrm{Im}\left[\sigma_\mathsmaller{-}(\omega)\right]$.}\label{fig:GeneralConductivityRefractive}
\end{figure}

\subsection{General behavior}\label{SubSec:General}
To give a general impression of the effect the electron-phonon interaction has on the magneto-optical conductivity and of the quantitative differences between substrates, Figs.~\ref{fig:GeneralConductivity}~(a) and~(b) show the conductivities $\mathrm{Re}\left[\sigma_\mathsmaller{xx}(\omega)\right]$ and $\mathrm{Im}\left[\sigma_\mathsmaller{xy}(\omega)\right]$ as fractions of the universal ac conductivity $\sigma_0=e^2/(4\hbar)$ at room temperature, $B=10$ T, a fixed chemical potential $\mu=0.2$ eV, and $\Gamma=5$ meV for graphene on several different substrates: Al$_2$O$_3$, hexagonal BN, HfO$_2$, SiC, SiO$_2$, and a non-polar substrate (where only the intrinsic graphene optical phonons affect the conductivity). For comparison, $\mathrm{Re}\left[\sigma_\mathsmaller{xx}(\omega)\right]$ in the absence of any phonons is also presented in Fig.~\ref{fig:GeneralConductivity}~(a). Moreover, the real parts of $\sigma_\mathsmaller{\pm}(\omega)=\sigma_\mathsmaller{xx}(\omega)\pm\i\sigma_\mathsmaller{xy}(\omega)$, describing the absorption of right- and left-handed circularly polarized light, respectively, are presented in Figs.~\ref{fig:GeneralConductivity}~(c) and~(d).

One can clearly see a pronounced absorption peak in Figs.~\ref{fig:GeneralConductivity}~(a)-(c), which corresponds to intraband transitions, the main contribution to which arises from $n=3\to4$ transitions. The interaction between electrons and phonons leads to a modification of the LLs (see Sec.~\ref{Sec:Intro}), which is reflected in a shift of the position of the intraband absorption peak to lower energies compared to the case where no phonons are considered. The magnitude of this polaronic shift is smallest for non-polar substrates, followed by that of the polar substrates BN and SiO$_2$. For substrates with a strong electron-SPP coupling such as Al$_2$O$_3$ or with low SPP frequencies such as HfO$_2$ the magnitude of the shift is larger. In the case of HfO$_2$ one can furthermore also discern several sidepeaks corresponding to phonon-assisted transitions [see Fig.~\ref{fig:GeneralConductivity}~(c)]. In addition to the absorption presented in Fig.~\ref{fig:GeneralConductivity}, the corresponding refractive components of the magneto-optical conductivities, which are also of experimental interest, are shown in Fig.~\ref{fig:GeneralConductivityRefractive}.

\begin{figure}[t]
\centering
\includegraphics*[width=8cm]{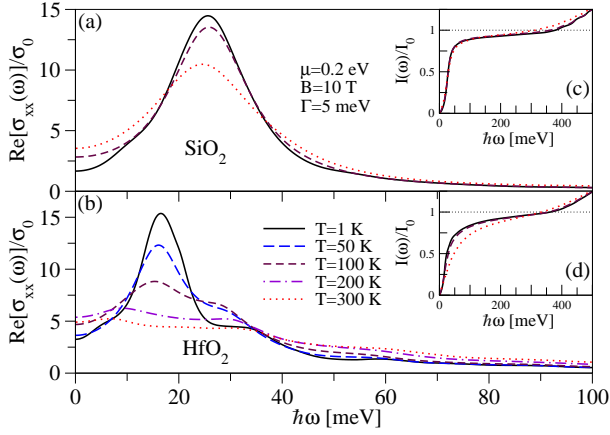}
\caption{(Color online) Calculated frequency dependence of $\mathrm{Re}\left[\sigma_\mathsmaller{xx}(\omega)\right]$ and of the corresponding spectral weight $I(\omega)$ for graphene on (a,c) SiO$_2$ and (b,d) HfO$_2$ substrates at different temperatures $T$.}\label{fig:ConductivityTempIntra}
\end{figure}

\begin{figure}[t]
\centering
\includegraphics*[width=8cm]{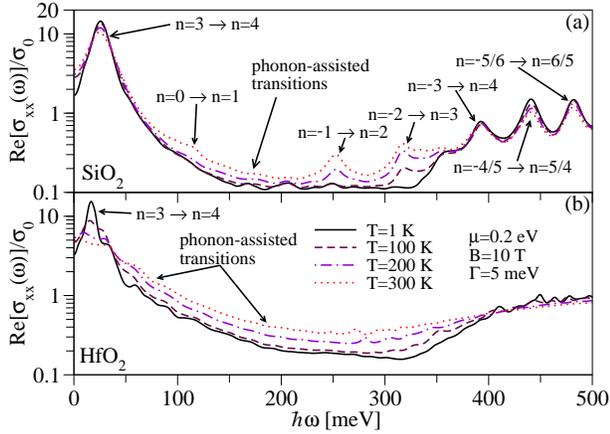}
\caption{(Color online) Calculated frequency dependence of $\mathrm{Re}\left[\sigma_\mathsmaller{xx}(\omega)\right]$ for graphene on (a) SiO$_2$ and (b) HfO$_2$ substrates at different temperatures $T$. The different intra- and interband transitions are labeled in the case of SiO$_2$.}\label{fig:ConductivityTempInter}
\end{figure}

\subsection{Temperature dependence}\label{SubSec:Temp}
The pronounced difference  at $T=300$ K described above between HfO$_2$ and other substrates is a consequence of the strong temperature dependence introduced by the SPPs. This is illustrated in Figs.~\ref{fig:ConductivityTempIntra} and~\ref{fig:ConductivityTempInter}, which compare the optical absorption $\mathrm{Re}\left[\sigma_\mathsmaller{xx}(\omega)\right]$ of SiO$_2$ with that of HfO$_2$ at different temperatures. As can be seen in Fig.~\ref{fig:ConductivityTempIntra}, which focuses on the intraband absorption, the intraband absorption peak (here: $n=3\to4$) becomes broader and loses spectral weight with increasing temperature.

There are two reasons for this loss of spectral weight, which happens for both SiO$_2$ and HfO$_2$: (i) As temperature increases, LLs close below the Fermi level are thermally depopulated, while LLs above the Fermi level are populated. Hence, new transitions (satisfying the selection rules) are possible into the now no longer fully occupied LLs slightly below $\mu$ or from the no longer completely empty LLs slightly above $\mu$. In the case of intraband transitions the new transitions are close to the main absorption peak, which leads to a broadening of the intraband absorption peak. (ii) With increasing temperature the number of phonons and thus the probability of electron-phonon scattering (leading also to a broadening of the absorption peaks) as well as of phonon-assisted transitions increases.

Comparing SiO$_2$ and HfO$_2$, one can discern that the main mechanism for the behavior of SiO$_2$ (up to room temperature) is (i), while HfO$_2$ is also strongly affected by (ii) for temperatures above approximately 200 K. Figures~\ref{fig:ConductivityTempIntra}~(c) and~(d) show the spectral weight
\begin{equation}\label{spectralweight}
I(\omega)=\int\limits_{0}^{\omega}\d\omega^\mathsmaller{\mathrm{\prime}}\mathrm{Re}\left[\sigma_\mathsmaller{xx}(\omega^\mathsmaller{\mathrm{\prime}})\right]
\end{equation}
normalized to the non-interacting spectral weight $I_0=2|\mu|\sigma_0/\hbar$ for of SiO$_2$ and HfO$_2$, respectively. They illustrate that the spectral weight lost at the main peak is mainly redistributed in the region between the intraband peak and the first interband peaks.

The mechanisms (i) and (ii) detailed above for the intraband transitions also apply to the interband transitions: In the case of SiO$_2$ [Fig.~\ref{fig:ConductivityTempInter}~(a)] one can clearly observe the emergence of new pronounced interband transitions due to the thermal depopulation of LLs below the Fermi level ($n=0\to1,-1\to2,-2\to3$) at higher temperatures, while phonon-assisted transitions also lead to an enhanced midgap absorption, that is, the absorption in the gap between intra- and interband transitions. For graphene on a HfO$_2$ substrate [Fig.~\ref{fig:ConductivityTempInter}~(b)] one cannot even observe pronounced interband transition peaks anymore at higher temperatures due to increased electron-phonon scattering. Likewise, the midgap absorption in HfO$_2$ rapidly increases from approximately $20\%$ at low temperatures to about $40\%$ at room temperature due to additional phonon-assisted transitions, whereas the increase of the midgap absorption from about $12-13\%$ to $15-16\%$ is much more modest in SiO$_2$.

\begin{figure}[t]
\centering
\includegraphics*[width=8cm]{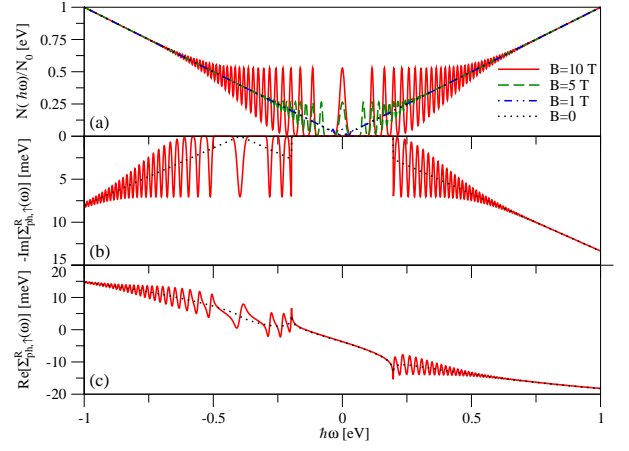}
\caption{(Color online) (a) Energy dependence of the spin-unpolarized DOS $N(\hbar\omega)$ for several different magnetic fields, where $N_0=2/(\pi\hbar^2v^2_\mathrm{\mathsmaller{F}})$. At high energies where the spacing between LLs becomes comparable or smaller than the lifetime broadening, $N(\hbar\omega)$ converges to the DOS of zero magnetic field. (b) Imaginary and (c) real parts of the self-energy $\Sigma^{\mathsmaller{\mathrm{R}}}_{\mathsmaller{\mathrm{ph},\uparrow}}\left(\omega\right)$ at $T=0$ and $\mu=0.2$ eV if only intrinsic graphene optical phonons at the $\Gamma$ point are taken into account.}\label{fig:DOS}
\end{figure}

\begin{figure}[t]
\centering
\includegraphics*[width=8cm]{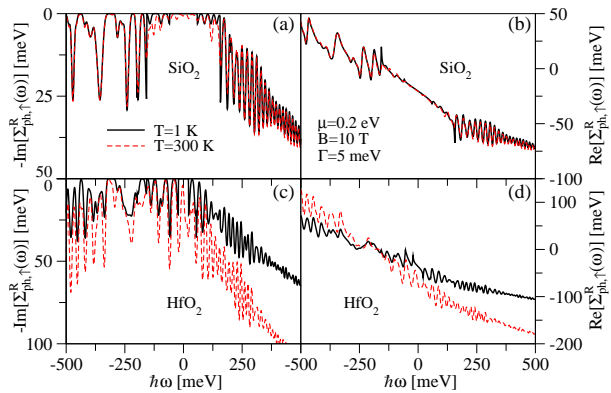}
\caption{(Color online) Calculated energy dependence of the self-energy $\Sigma^{\mathsmaller{\mathrm{R}}}_{\mathsmaller{\mathrm{ph},\uparrow}}\left(\omega\right)$ for graphene on (a,b) SiO$_2$ and (c,d) HfO$_2$ substrates at $T=1$ K and $T=300$ K.}\label{fig:SelfEnergy}
\end{figure}

\subsection{Temperature dependence of the self-energy}\label{SubSec:TempSE}
While mechanism (i) is a consequence of Fermi statistics and described by the factor $n_\mathsmaller{\mathrm{FD}}\left(\hbar\omega'\right)-n_\mathsmaller{\mathrm{FD}}\left(\hbar\omega''\right)$ in Eqs.~(\ref{conductivity_xx}) and~(\ref{conductivity_xy}), the effect of phonons, that is, mechanism (ii), is described by the self-energy~(\ref{self_energy_phonons}). It is this self-energy due to phonons, $\Sigma^{\mathsmaller{\mathrm{R}}}_{\mathsmaller{\mathrm{ph},s}}\left(\omega\right)$, from which the qualitative difference between SiO$_2$ and HfO$_2$ at room temperature arises.

Since the self-energy is calculated from the spin-polarized DOS, which in turn can be expressed by the spin-unpolarized DOS $N(\epsilon)$, Fig.~\ref{fig:DOS}~(a) shows $N(\epsilon)$ at several different magnetic fields for the reader's orientation. The imaginary part of $\Sigma^{\mathsmaller{\mathrm{R}}}_{\mathsmaller{\mathrm{ph},s}}\left(\omega\right)$ describing electron-phonon scattering is essentially given by contributions from the DOS centered around $-\mu+sg\mu_\mathsmaller{B}B/2\pm\hbar\omega_\mathsmaller{\Lambda}$ and multiplied by the thermal factor $n_\mathsmaller{\mathrm{BE}}(\hbar\omega_\mathsmaller{\Lambda})+n_\mathsmaller{\mathrm{FD}}(\hbar\omega_\mathsmaller{\Lambda}\mp\hbar\omega)$ for each individual phonon branch $\Lambda$. At $T=0$ these thermal factors are just Heaviside step functions as $n_\mathsmaller{\mathrm{BE}}(\hbar\omega_\mathsmaller{\Lambda})$ vanishes, resulting in a gap of width $2\hbar\omega_\mathsmaller{\Lambda}$ centered around $\hbar\omega=0$ for each phonon branch. This is illustrated in Fig.~\ref{fig:DOS}~(b), where the contribution from $\Lambda=\Gamma$ to $\mathrm{Im}\left[\Sigma^{\mathsmaller{\mathrm{R}}}_{\mathsmaller{\mathrm{ph},\uparrow}}\left(\omega\right)\right]$ at $T=0$ and $\mu=0.2$ eV is presented, with the corresponding real part displayed in Fig.~\ref{fig:DOS}~(c). Since Zeeman splitting for $g=2$ at $B=10$ T is small compared to the LL broadening, the spin-degree of freedom does not affect the self-energy noticeably and the self-energies of spin-up and spin-down electrons are almost identical.

The total self-energy due to phonons is then given by the sum of the individual contributions from each branch $\Lambda$ and shown in Fig.~\ref{fig:SelfEnergy} for spin-up electrons at $B=10$ T, $\mu=0.2$ eV, and $T=1$ K as well $T=300$ K. At finite temperatures the discontinuities at $\omega=\pm\omega_\mathsmaller{\Lambda}$ disappear and are thermally broadened with increasing temperature. Moreover, there is now also a finite contribution from $n_\mathsmaller{\mathrm{BE}}(\hbar\omega_\mathsmaller{\Lambda})$ which grows with increasing temperature. For phonons with relatively high frequencies $\omega_\mathsmaller{\Lambda}$, such as in Figs.~\ref{fig:SelfEnergy}~(a) and~(b), this contribution results in the emergence of peaks inside the region $\left[-\omega_\mathsmaller{\Lambda},\omega_\mathsmaller{\Lambda}\right]$, but does not significantly affect the high-frequency behavior at room temperature dominated by $n_\mathsmaller{\mathrm{FD}}(-|\hbar\omega|)\approx1$ as $n_\mathsmaller{\mathrm{BE}}(\hbar\omega_\mathsmaller{\Lambda})\ll1$. In HfO$_2$ the situation is quite different at room temperature, which is entirely due to the contribution from the SO$_1$ phonon: Because of the low frequency $\hbar\omega_\mathsmaller{\mathrm{SO}_1}=21.6$ meV, which corresponds to $n_\mathsmaller{\mathrm{BE}}(\hbar\omega_\mathsmaller{SO_1})\approx0.77$, the self-energy at high frequencies is strongly enhanced compared to its low-temperature values [see Figs.~\ref{fig:SelfEnergy}~(c) and~(d)]. As a consequence of the significantly enhanced imaginary part of the self-energy the transition peaks in the optical conductivity of graphene on HfO$_2$ are strongly broadened, while the increased real part of the self-energy results in a growing polaronic shift to lower energies [compare Fig.~\ref{fig:ConductivityTempIntra}~(b)].

The behavior of the self-energy of SiO$_2$ and HfO$_2$ described above reflects the fact that at room temperature there are many more SO$_1$ phonons available for scattering with electrons or phonon-assisted optical transitions in graphene on HfO$_2$ than there are in graphene on SiO$_2$. In this way, the stronger effect of polar substrates with low SPP frequencies, such as HfO$_2$, compared to polar substrates with higher SPP frequencies or non-polar substrates can be understood intuitively. Finally, we note that, if the temperature was increased further, graphene on SiO$_2$ or other polar substrates would eventually also show a behavior similar to that of HfO$_2$.

\begin{figure}[t]
\centering
\includegraphics*[width=8cm]{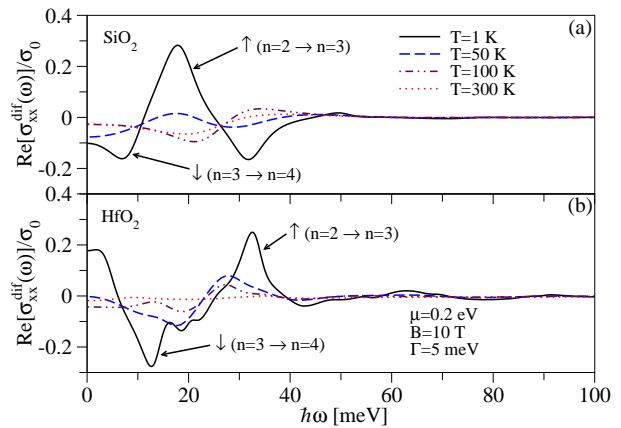}
\caption{(Color online) Calculated frequency dependence of the spin-polarized magneto-optical conductivity $\mathrm{Re}\left[\sigma^{\mathsmaller{\mathrm{dif}}}_{\mathsmaller{xx}}(\omega)\right]$ for graphene on (a) SiO$_2$ and (b) HfO$_2$ substrates at different temperatures $T$.}\label{fig:SpinConductivityTempIntra}
\end{figure}

\subsection{Spin-polarized absorption}\label{SubSec:Spin}
As mentioned above, the spin-degree of freedom does not play a crucial role for magneto-optical absorption experiments because Zeeman splitting is small at experimentally relevant magnetic fields, a fact not only reflected in $\sigma_{\mathsmaller{xx}}(\omega)$, but also in the self-energy seen above. Nevertheless, there can be some spin effects---albeit small, namely at a level of $1\%$ of the total absorption---as demonstrated by Fig.~\ref{fig:SpinConductivityTempIntra}. There the difference between the (intraband) absorption by spin-up and by spin-down electrons is shown, that is, $\mathrm{Re}\left[\sigma^{\mathsmaller{\mathrm{dif}}}_{\mathsmaller{xx}}(\omega)\right]=\mathrm{Re}\left[\sigma^{\uparrow}_{\mathsmaller{xx}}(\omega)\right]-\mathrm{Re}\left[\sigma^{\downarrow}_{\mathsmaller{xx}}(\omega)\right]$, where $\sigma^{\uparrow/\downarrow}_{\mathsmaller{xx}}(\omega)$ is given by Eq.~(\ref{conductivity_xx}) if the spin-degree of freedom is fixed to $s=\uparrow/\downarrow$ and the summation over $s$ is omitted. The substrates are again chosen to be SiO$_2$ and HfO$_2$ with $\mu=0.2$ eV, $B=10$ T, $\Gamma=5$ meV, and several different temperatures, that is, the same choice of parameters as in Figs.~\ref{fig:GeneralConductivity}-\ref{fig:ConductivityTempInter}.

As can be seen in Fig.~\ref{fig:SpinConductivityTempIntra}, there is a slight difference between the amount of absorption by spin-up and spin-down electrons in the intraband transition peak seen in Figs.~\ref{fig:GeneralConductivity}-\ref{fig:ConductivityTempInter}. This is primarily a consequence of the fact that the $n=3$ LL for spin-up is extremely close to the Fermi level $\mu=0.2$ eV and (due to disorder) slightly depopulated. Thus, even at low temperatures there is a slightly enhanced probability of transitions $n=2\to3$ for spin-up electrons compared to spin-down electrons. Consequently, the probability of transitions $n=3\to4$ at low temperatures is slightly higher for spin-down electrons than for spin-up electrons. The most pronounced peaks in Fig.~\ref{fig:SpinConductivityTempIntra} can be attributed to this imbalance between optical transitions. In addition to this essentially single-particle effect there is also a substructure arising from the electron-phonon interaction as can particularly be seen in Fig.~\ref{fig:SpinConductivityTempIntra}~(b). At higher temperatures the spin-polarized conductivity $\sigma^{\mathsmaller{\mathrm{dif}}}_{\mathsmaller{xx}}(\omega)$ quickly decreases as $k_{\mathsmaller{\mathrm{B}}}T$ exceeds the value of the Zeeman splitting.

\begin{figure}[t]
\centering
\includegraphics*[width=8cm]{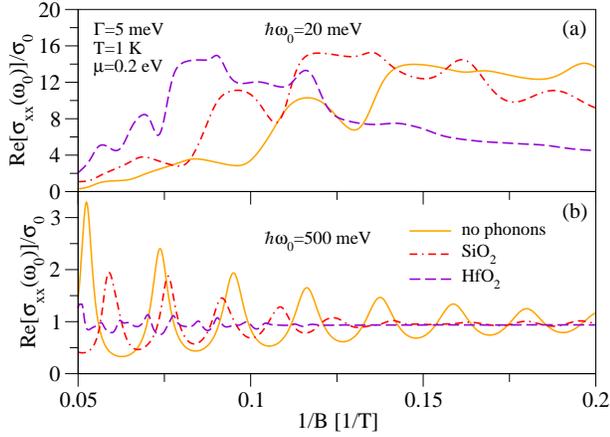}
\caption{(Color online) Calculated magnetic field dependence of $\mathrm{Re}\left[\sigma_\mathsmaller{xx}(\omega_0)\right]$ for graphene on SiO$_2$ and HfO$_2$ substrates at (a) $\hbar\omega_0=20$ meV and (b) $\hbar\omega_0=500$ meV. For comparison, $\mathrm{Re}\left[\sigma_\mathsmaller{xx}(\omega_0)\right]$ in the absence of any phonons is also presented.}\label{fig:Conductivity1oB}
\end{figure}

\subsection{Dependence on the magnetic field and chemical potential}\label{SubSec:MagField}
Having investigated the dependence of the absorption on temperature, we now briefly discuss how changing the magnetic field or chemical potential affects the magneto-optical conductivity. Figure~\ref{fig:Conductivity1oB} depicts $\mathrm{Re}\left[\sigma_\mathsmaller{xx}(\omega_0)\right]$ at $\hbar\omega_0=20$ meV and $\hbar\omega_0=500$ meV as a function of the inverse magnetic field for graphene on SiO$_2$ and on HfO$_2$ with $\mu=0.2$ eV, $T=1$ K, and $\Gamma=5$ meV. For comparison, we also show the absorption obtained from the single-particle picture, that is, in the absence of any phonons. As before, the qualitative behavior can be understood from the single-particle picture: With increasing $B$ the spacing between LLs increases and different intraband and interband transitions---governed by the selection rules and the position of the Fermi level with respect to the LLs---become possible. This leads to the behavior depicted in Fig.~\ref{fig:Conductivity1oB}~(a) for intraband transitions monitored at low frequencies such as $\hbar\omega_0=20$ meV and the oscillatory behavior of interband transitions monitored at frequencies above $2|\mu|$ and illustrated in Fig.~\ref{fig:Conductivity1oB}~(b). One can also clearly observe a polaronic shift due to the interaction between electrons and phonons for both intra- and interband transitions. Moreover, with decreasing magnetic field the amplitudes of the interband transition oscillations around $\sigma_0$ in Fig.~\ref{fig:Conductivity1oB}~(b) decrease as the ratio between the LL spacing and the broadening $\Gamma$ is diminished.

\begin{figure}[t]
\centering
\includegraphics*[width=8cm]{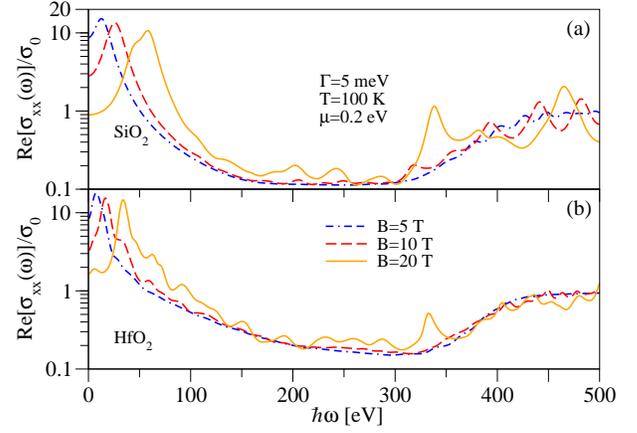}
\caption{(Color online) Calculated frequency dependence of $\mathrm{Re}\left[\sigma_\mathsmaller{xx}(\omega)\right]$ for graphene on (a) SiO$_2$ and (b) HfO$_2$ substrates with different magnetic fields $B$.}\label{fig:ConductivityMagField}
\end{figure}

\begin{figure}[t]
\centering
\includegraphics*[width=8cm]{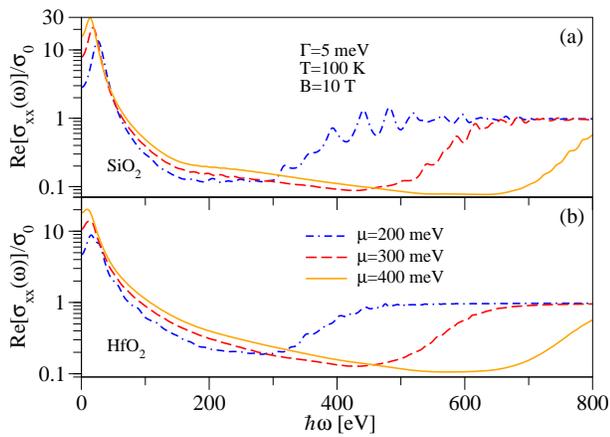}
\caption{(Color online) Calculated frequency dependence of $\mathrm{Re}\left[\sigma_\mathsmaller{xx}(\omega)\right]$ for graphene on (a) SiO$_2$ and (b) HfO$_2$ substrates at different chemical potentials $\mu$.}\label{fig:ConductivityChemPot}
\end{figure}

The statements given above are corroborated in Fig.~\ref{fig:ConductivityMagField}, which displays the frequency dependence of $\mathrm{Re}\left[\sigma_\mathsmaller{xx}(\omega)\right]$ for graphene on SiO$_2$ and HfO$_2$ substrates at $T=100$ K, $\mu=0.2$ eV, $\Gamma=5$ meV, and different magnetic fields. The major trends that can be seen here are that with increasing magnetic field the intraband transition peak moves to higher energies, while the amplitudes of interband transition peaks and of phonon-assisted peaks increase. A somewhat opposite behavior can be observed in Fig.~\ref{fig:ConductivityChemPot}, which shows the dependence of $\mathrm{Re}\left[\sigma_\mathsmaller{xx}(\omega)\right]$ on the chemical potential at $T=100$ K, $B=10$ T, $\Gamma=5$ meV. As the chemical potential increases, the intraband transition peak moves to lower energies and the amplitudes of the oscillations for interband transitions decrease. Furthermore, the onset of interband transitions in the vicinity of $2|\mu|$ can also be clearly followed in Fig.~\ref{fig:ConductivityChemPot}. 

Both the behavior of the intraband transition peak with increasing magnetic field and with increasing chemical potential can be qualitatively explained in the single-particle picture as originating from the LL spectrum in the vicinity of the Fermi level: With increasing magnetic field the LL spacing increases giving rise to a higher energy of the intraband transition. On the other hand, for a fixed magnetic field the LL spacing near the Fermi level decreases if the absolute value of the chemical potential is increased and thus situated in a denser region of the LL spectrum. Consequently, the energy of the intraband transition is decreased. Likewise, the decrease of the amplitudes of the interband peaks with increasing chemical potential $\mu$, decreasing magnetic fields $B$, or high photon energies $\hbar\omega$ can be interpreted as arising from optical transitions between increasingly denser regions of the LL spectrum, where the energy difference between different transitions is small compared to the broadening due to scattering.

While the main trends in Figs.~\ref{fig:ConductivityMagField} and \ref{fig:ConductivityChemPot} can thus be readily understood from the single-particle picture, the electron-phonon coupling has still a profound impact and is needed to explain features such as polaronic shifts, the emergence of phonon sidebands, and enhanced broadening of the transition peaks, particularly striking at high photon energies $\hbar\omega$ or higher temperatures, due to electron-phonon scattering.

\begin{figure}[t]
\centering
\includegraphics*[width=8cm]{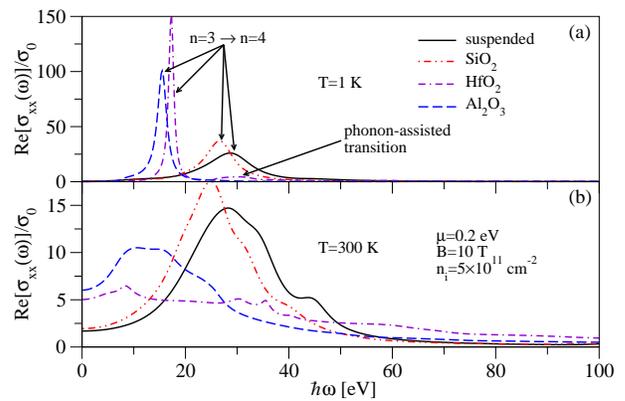}
\caption{(Color online) Calculated frequency dependence of $\mathrm{Re}\left[\sigma_\mathsmaller{xx}(\omega)\right]$ for graphene on several different substrates at (a) $T=1$ K and (b) $T=300$ K in the presence of Coulomb impurity scattering.}\label{fig:GeneralConductivityNimp}
\end{figure}

\subsection{Scattering at Coulomb impurities}\label{SubSec:Coulomb}
So far, our model has accounted for scattering other than electron-optical phonon/SPP scattering by the addition of a constant scattering rate $\Gamma$ in the total self-energy~\ref{total_self_energy}. However, in a more realistic model this scattering rate would also be dependent on the energy and LL of the state considered. To model this behavior at least partially, we assume that scattering is dominated by scattering at charged impurities and replace $\Gamma$ by $\hbar/[2\tau(\epsilon_k=\epsilon_{s}(n))]$, where for simplicity we use the transport scattering time $\tau(\epsilon_k)$ due to Coulomb impurities for an electron with energy $\epsilon_k$ as calculated in Ref.~\onlinecite{Hwang2009:PRB} for $B=0$ and evaluate it at the energy corresponding to the LL considered. Since Coulomb impurity scattering also depends on the dielectric environment, the scattering rate $\hbar/[2\tau(\epsilon_k=\epsilon_{s}(n))]$ is also affected by the choice of the substrate.

This can be seen in Fig.~\ref{fig:GeneralConductivityNimp}, which shows $\mathrm{Re}\left[\sigma_\mathsmaller{xx}(\omega)\right]$ at $T=1$ K and $T=300$ K for suspended graphene ($\kappa=1$) and for graphene on several different substrates with an impurity concentration of $n_\mathsmaller{\mathrm{i}}=5\times10^{11}$ cm$^{-2}$, $\mu=0.2$ eV, and $B=10$ T. At low temperatures [see Fig.~\ref{fig:GeneralConductivityNimp}~(a)] the broadening of the intraband transition peaks is mainly determined by the Coulomb impurity scattering and markedly different for different substrates. By comparing Fig.~\ref{fig:GeneralConductivityNimp}~(a) with the low-temperature behavior in Fig.~\ref{fig:ConductivityTempIntra}, one can also conclude that the actual intraband peak positions themselves are not noticeably affected by the inclusion of a LL-dependent scattering rate. If the temperature is increased, electron-phonon scattering becomes the dominant scattering mechanism and thus the quantitative differences between the results obtained for constant and LL-dependent scattering rates are no longer as pronounced as at low temperatures as can be deduced from a comparison of Figs.~\ref{fig:GeneralConductivity}~(a) and~\ref{fig:GeneralConductivityNimp}~(b).

\subsection{Comparison with experiments}\label{SubSec:Exp}
As mentioned in Sec.~\ref{Sec:Intro}, several distinct absorption peaks can be seen in experiments.\cite{Sadowski2006:PRL,Jiang2007:PRL,Sadowski2007:SolidStateComm,Deacon2007:PRB,Henriksen2010:PRL,Orlita2010:SST,Crassee2011:PRB,Orlita2011:PRL} The positions of those peaks, however, are not exactly where the single-particle picture predicts they are, and the experimental data is usually fitted to the single-particle picture using a renormalized Fermi velocity. This procedure results in different values of the fitted Fermi velocity in different experiments and in the case of Refs.~\onlinecite{Jiang2007:PRL,Crassee2011:PRB} even for different LL transitions. The experimental values for the fitted Fermi velocity range from around $\tilde{c}\approx1.03\times10^6$ m/s to $\tilde{c}\approx1.12\times10^6$ m/s. Those values are higher than the bare Fermi velocity, indicating that the transition peaks are shifted to higher energies.

This shift to higher energies is attributed to the electron-electron interaction, which---in contrast to a standard two-dimensional electron gas\cite{Kohn1961:PhysRev}---is argued to play an important role in the LL spectroscopy of graphene.\cite{Jiang2007:PRL,Iyengar2007:PRB,Bychkov2008:PRB,Henriksen2010:PRL} However, while electron-electron interaction predicts the correct sign of the shift (that is, to higher energies), it also predicts a very strong correction of approximately $30\%$, not seen in experiments (see above).

As described in the previous sections as well as in Ref.~\onlinecite{Pound2012:PRB}, the electron-phonon interaction also leads to a shift of the transition peaks, albeit to lower energies. Thus, it stands to reason that the strong effect due to the electron-electron interaction is somewhat compensated by the electron-SPP coupling. Most magneto-optical experiments up until now have been conducted on SiO$_2$ or SiC substrates, where our calculations suggest a shift to lower energies corresponding to a correction of approximately $15-20\%$ compared to the position of the bare transition peak at low temperatures, $\mu=0.2$ eV, and $B=10$ T [see Fig.~\ref{fig:GeneralConductivityNimp}~(a)]. Experiments on substrates with strong electron-phonon coupling, such as Al$_2$O$_3$, or with low SPP frequencies, such as HfO$_2$, could further clarify the role played by SPPs. Our calculations suggest that for such substrates the effect of the electron-phonon coupling reducing the electron-electron-interaction-induced shift is even more pronounced and might possibly even completely compensate for it.

To give a crude estimate of the combined impact of the electron-phonon and electron-electron interaction, we have calculated the optical absorption at $T=1$ K, $\mu=0.2$ eV, and $B=18$ T using the procedure outlined in Sec.~\ref{Sec:Model}, where we have replaced the Fermi velocity $v_\mathrm{\mathsmaller{F}}\approx 10^6$ m/s by a higher value of $1.30\times10^6$ m/s to reflect the effect of the Coulomb interaction. In this way we find that the positions of the intraband transition peaks for graphene on SiO$_2$ and SiC substrates are shifted to higher energies compared to the bare case (that is, in the absence of phonons and with a Fermi velocity of $v_\mathrm{\mathsmaller{F}}\approx 10^6$ m/s) and obtain fitted Fermi velocities of $\tilde{c}\approx1.07\times10^6$ m/s for SiC and of $\tilde{c}\approx1.11\times10^6$ m/s for SiO$_2$. While these values are significantly closer to experimental data, we stress that these values are only simple estimates and that a more detailed theory is required to treat the renormalization of the Fermi velocity consistently.\cite{Park2007:PRL,Park2009:NL,Elias2011:NatPhys}

Due to the small size of the Zeeman splitting for $g=2$ compared to the electron-phonon or electron-impurity scattering, there is only a weak spin dependence of the magneto-optical conductivity. However, substantially larger $g$-factors have been reported for hydrogenated graphene,\cite{McCreary2012:PRL} making an enhancement of the spin effects briefly discussed in Sec.~\ref{SubSec:Spin} appear more feasible, although in that case there is also enhanced scattering due to the presence of hydrogen adatoms, somewhat compensating for the effect of an increased g-factor.

\section{Conclusions}\label{Sec:Conclusions}
In this work we have investigated how the magneto-optical conductivity of (doped) monolayer graphene is affected by the substrate the graphene layer is placed upon. Here our particular focus has been on the effects of SPPs. Our calculations suggest that polaronic shifts of the intra- and interband absorption peaks can be significantly enhanced for substrates with strong electron-SPP coupling as compared to those for non-polar substrates, where only intrinsic graphene optical phonons contribute. Moreover, electron-phonon scattering and phonon-assisted transitions result in a broadening and loss of spectral weight at the transition peaks. The strength of these processes is strongly temperature-dependent and with high temperatures the magneto-optical conductivity becomes increasingly affected by polar substrates. This is especially true for polar substrates with small SPP energies such as HfO$_2$, where many phonons are available for scattering and phonon-assisted transitions and most of the spectral weight has been transferred away from the main absorption peaks already at room temperature. 

Furthermore, we have also briefly studied the effect of LL-dependent scattering rates modeling Coulomb impurity scattering. While the qualitative picture of the impact of optical and SPPs on the magneto-optical conductivity, outlined above, remains unaffected by the inclusion of a LL-dependent broadening, it can play a profound role in determining the lineshape of the absorption peaks, especially at low temperatures, where impurity scattering dominates.

\acknowledgments
We gratefully acknowledge John Cerne from the University at Buffalo for stimulating discussions. This work was supported by U.S. ONR, NSF-NRI NEB 2020, and SRC. J.F. acknowledges support from DFG Grant No. GRK 1570.

\appendix

\section{Total self-energy and Kubo formulas for the magneto-optical conductivity}\label{Sec:Appendix}
In order to calculate the self-energy~(\ref{self_energy_phonons}) due to phonons, the spin-polarized DOS is needed. Using a constant Gaussian broadening\footnote{For convenience, we use Gaussian instead of Lorentzian broadening to calculate the DOS here. However, we find that the results presented in this manuscript are not significantly affected by using Lorentzian broadening.} of width $\Gamma$ for each state to describe scattering other than scattering with optical or SPPs, such as scattering at charged impurities, on a phenomenological level and employing Poisson's summation formula, we can write the spin-polarized DOS for the spectrum given in Eq.~(\ref{electronic_single_particle_spectrum}) as $N_s(\epsilon)=N(\epsilon-sg\mu_\mathsmaller{B}B/2)/2$, where
\begin{equation}\label{DOS}
\begin{aligned}
N&(\epsilon\neq0)=\frac{2|\epsilon|}{\pi\hbar^2v^2_\mathrm{\mathsmaller{F}}}\Bigg\{1+\\
&2\sum\limits_{m=1}^{\infty}\exp\left[-\left(\frac{\sqrt{2}\pi m\Gamma l^2_\mathsmaller{B}\epsilon}{\hbar^2v^2_\mathrm{\mathsmaller{F}}}\right)^2\right]\cos\left(\frac{\pi ml^2_\mathsmaller{B}\epsilon^2}{\hbar^2v^2_\mathrm{\mathsmaller{F}}}\right)\Bigg\},\\
N&(\epsilon=0)=\frac{4\Gamma}{\pi\hbar^2v^2_\mathrm{\mathsmaller{F}}\sqrt{2\pi}}\frac{\hbar^2v^2_\mathrm{\mathsmaller{F}}/(2\Gamma^2l^2_\mathsmaller{B})}{\tanh\left[\hbar^2v^2_\mathrm{\mathsmaller{F}}/(2\Gamma^2l^2_\mathsmaller{B})\right]}.
\end{aligned}
\end{equation}
Thus, Eqs.~(\ref{self_energy_phonons})-(\ref{Froehlich_SOi}) and~(\ref{DOS}) allow us to calculate the electronic self-energy due to phonons. The real part of Eq.~(\ref{self_energy_phonons}) is related to the polaron formation, while its imaginary part describes the scattering between electrons and phonons.

Finally, the effect of non-phonon-related scattering is also included in the electronic self-energy by adding the constant scattering rate $\Gamma$. Then the total self-energy can be obtained as
\begin{equation}\label{total_self_energy}
\begin{aligned}
\Sigma&^{\mathsmaller{\mathrm{R}}}_{s}\left(\omega\right)=-\i\Gamma+\sum\limits_\Lambda\frac{A_\mathsmaller{\Lambda}}{2}\int\limits_{-\infty}^\infty\d\omega'N(\hbar\omega'+\mu-sg\mu_\mathsmaller{B}B/2)\\
&\times\bigg[\frac{n_\mathsmaller{\mathrm{BE}}\left(\hbar\omega_\mathsmaller{\Lambda}\right)+n_\mathsmaller{\mathrm{FD}}\left(-\hbar\omega'\right)}{\omega-\omega'-\omega_\mathsmaller{\Lambda}+\i0^\mathsmaller{+}}+\frac{n_\mathsmaller{\mathrm{BE}}\left(\hbar\omega_\mathsmaller{\Lambda}\right)+n_\mathsmaller{\mathrm{FD}}\left(\hbar\omega'\right)}{\omega-\omega'+\omega_\mathsmaller{\Lambda}+\i0^\mathsmaller{+}}\bigg],
\end{aligned}
\end{equation}
which can in turn be used to extract the electronic Green's function and the corresponding spectral function
\begin{equation}\label{electron_spectral_function}
\mathcal{A}_{n,s}(\omega)=-2\,\mathrm{Im}\left\{\frac{1}{\omega+\i0^\mathsmaller{+}-\left[\epsilon_{s}(n)-\mu+\Sigma^{\mathsmaller{\mathrm{R}}}_{s}(\omega)\right]/\hbar}\right\}.
\end{equation}

The electronic spectral function~(\ref{electron_spectral_function}) can then be used to calculate the magneto-optical conductivities as\footnote{Those formulas are straight-forward extensions of the formulas found in V. P. Gusynin, S. G. Sharapov, and J. P. Carbotte, J. Phys.: Condens. Matter \textbf{19}, 026222 (2007) and in Ref.~\onlinecite{Pound2012:PRB} to account for the spin-degree of freedom.}
\begin{equation}\label{conductivity_xx}
\begin{aligned}
\sigma_{xx}\left(\omega\right)=\frac{\i\sigma_0v^2_\mathrm{\mathsmaller{F}}}{4\pi^3\omega l^2_\mathsmaller{B}}&\sum\limits_{s,n=0}^\infty\int\limits_{-\infty}^{\infty}\d\omega'\d\omega''\frac{n_\mathsmaller{\mathrm{FD}}\left(\hbar\omega'\right)-n_\mathsmaller{\mathrm{FD}}\left(\hbar\omega''\right)}{\omega'-\omega''+\omega+\i0^\mathsmaller{+}}\\
&\times\left[\Psi^s_{n,n+1}(\omega',\omega'')+\Psi^s_{n+1,n}(\omega',\omega'')\right]
\end{aligned}
\end{equation}
and
\begin{equation}\label{conductivity_xy}
\begin{aligned}
\sigma_{xy}\left(\omega\right)=\frac{\sigma_0v^2_\mathrm{\mathsmaller{F}}}{4\pi^3\omega l^2_\mathsmaller{B}}&\sum\limits_{s,n=0}^\infty\int\limits_{-\infty}^{\infty}\d\omega'\d\omega''\frac{n_\mathsmaller{\mathrm{FD}}\left(\hbar\omega'\right)-n_\mathsmaller{\mathrm{FD}}\left(\hbar\omega''\right)}{\omega'-\omega''+\omega+\i0^\mathsmaller{+}}\\
&\times\left[\Psi^s_{n,n+1}(\omega',\omega'')-\Psi^s_{n+1,n}(\omega',\omega'')\right].
\end{aligned}
\end{equation}
In the derivation of the Kubo formulas~(\ref{conductivity_xx}) and~(\ref{conductivity_xy}) vertex corrections have been ignored. Moreover, both formulas include the universal ac conductivity $\sigma_0=e^2/(4\hbar)$ as well as the auxiliary function
\begin{equation}\label{auxiliary_function}
\begin{aligned}
\Psi^s_{m,n}(\omega',\omega'')=&\left[\mathcal{A}_{m,s}(\omega')+\mathcal{A}_{-m,s}(\omega')\right]\\
&\times\left[\mathcal{A}_{n,s}(\omega'')+\mathcal{A}_{-n,s}(\omega'')\right],
\end{aligned}
\end{equation}
which reflects the fact that (in the dipole approximation) only spin-conserving optical transitions from a given LL $|n|$ to LLs $|n\pm1|$ are permitted. Here we are primarily interested in the absorption, that is, essentially in $\mathrm{Re}\left[\sigma_\mathsmaller{xx}(\omega)\right]$. In the following, we will thus use Eqs.~(\ref{total_self_energy})-(\ref{auxiliary_function}) to calculate the real part of $\sigma_{xx}(\omega)=\sigma_{yy}(\omega)$ as well as the imaginary part of $\sigma_{xy}(\omega)=-\sigma_{yx}(\omega)$, where the $\omega''$ integration can be preformed using the Dirac-$\delta$ function arising from the denominator, while the remaining integral is computed numerically.\footnote{Our numerical integrations over $\omega$ have been conducted on grids with $\Delta(\hbar\omega)=0.1$ meV.} The imaginary part of $\sigma_{xx}(\omega)$ and the real part of $\sigma_{xy}(\omega)$, determining the refractive index in the graphene plane, can then be determined using the Kramers-Kronig relations.

\bibliographystyle{apsrev}

\end{document}